\documentclass[%
 aip,
 amsmath,amssymb,
reprint, 
]{revtex4-1}

\usepackage{graphicx}
\usepackage{dcolumn}
\usepackage{bm}

\usepackage[utf8]{inputenc}
\usepackage[T1]{fontenc}
\usepackage{mathptmx}
\usepackage{etoolbox}
\usepackage{import}
\usepackage{graphicx}
\usepackage{amsmath,amssymb,epstopdf}
\usepackage{dcolumn}
\usepackage{bm}
\usepackage{hyperref}
\usepackage{longtable}
\usepackage{booktabs}
\usepackage{multirow}
\usepackage{siunitx}
\usepackage{enumitem}

\newcommand{\pd}[2]{\frac{\partial{#1}}{\partial{#2}}}
\newcommand{\pdd}[2]{\frac{\partial^2{#1}}{\partial{#2}^2}}

\newcommand{\bsym}[1]{\boldsymbol{#1}}
\newcommand{\bu}{\bsym{u}}
\newcommand{\bx}{\bsym{x}}

\newcommand{\DT}{\Delta T}

\newcommand{\delsq}{\nabla^{2} }

\newcommand{\Gr}{G}
\newcommand{\Fr}{Fr}

\newcommand{\Ta}{Ta}
\newcommand{\kc}{k_c}
\newcommand{\bbR}{\mathbb{R}}
\newcommand{\udotgrad}[1]{u\pd {#1}{r} + S \frac{v}{r} \pd{#1}{\phi} + w \pd{#1}{z}  }
\newcommand{\NA}{\text{stable}}
\newcommand{\ub}{u_b}
\newcommand{\vb}{v_b}
\newcommand{\wb}{w_b}
\newcommand{\pb}{p_b}
\newcommand{\Tb}{T_b}
\newcommand{\Lapr} {\delsq u - S\frac{2}{r^{2}} \pd{v}{\phi} - \frac{u}{r^{2}}}
\newcommand{\Lapphi} {S \delsq v + \frac{2}{r^{2}}\pd{u}{\phi} - S\frac{v}{r^{2}}}

\newcommand{\aghoredit}[1]{#1}

\makeatletter
\def\@email#1#2{%
 \endgroup
 \patchcmd{\titleblock@produce}
  {\frontmatter@RRAPformat}
  {\frontmatter@RRAPformat{\produce@RRAP{*#1\href{mailto:#2}{#2}}}\frontmatter@RRAPformat}
  {}{}
}%
\makeatother
\begin{document}

\preprint{AIP/123-QED}

\title[]{Effect of outer cylinder rotation on the radially heated Taylor-Couette flow}
\author{Pratik Aghor}
 \email{$^1$pratikprashant.aghor@unh.edu $^2$fmohammad@bnl.gov}
 \affiliation{Integrated Applied Mathematics Program, University of New Hampshire, Durham, New Hampshire (US)}
\author{Mohammad Atif}%
\affiliation{%
 Brookhaven National Laboratory, Upton, New York (US)
}%

\date{\today}

\begin{abstract}
A Taylor--Couette setup with radial heating is considered where a Boussinesq fluid is sheared in the annular region between two concentric, independently rotating cylinders maintained at different temperatures.
Linear stability analysis is performed to determine the Taylor number for the onset of instability. 
Two radius ratios \aghoredit{corresponding to wide and thin gaps with several rotation rate ratios are considered}.
The rotation of the outer cylinder is found to have a general stabilizing effect on the stability threshold as compared to pure inner--cylinder rotation, with a few exceptions.
\aghoredit{
The radial heating sets up an axial flow which breaks the reflection symmetry of isothermal Taylor--Couette flow in the axial coordinate.
This symmetry breaking separates linear stability thresholds and we find fastest growing modes with both positive and negative azimuthal numbers for different parameters. 
Another important finding of the current study is the discovery of unstable modes in the Rayleigh-stable regime.
}
Furthermore, closed disconnected neutral curves (CDNCs) are observed for both wide and thin gaps \aghoredit{which} can separate from or merge into open neutral--stability curves. 
Alternatively, CDNCs can also morph into open neutral stability curves as the rotation rate ratio is changed. 
CDNCs are observed to be sensitive to changes in control parameters and their appearance/disappearance is shown to induce discontinuous jumps in the critical Taylor number. 
For both wide and thin gaps, the fastest--growing modes found in the pure corotation case are shown to have their origins in the instability islands at smaller values of rotation rate ratios. 
\end{abstract}

\maketitle

\section{Introduction}
\label{sec:intro}
\begin{figure*} 
 \begin{center}
  \includegraphics[width = 0.98\textwidth]{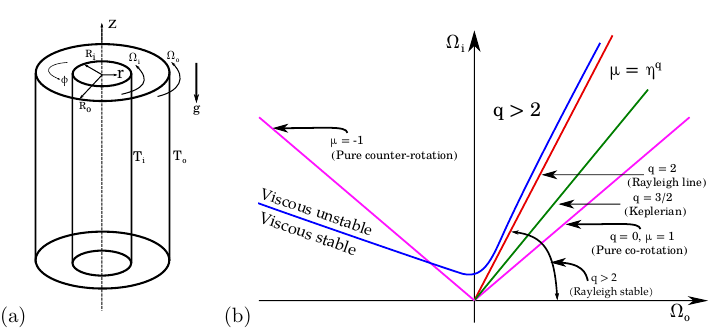}
\end{center}
\caption{\aghoredit{(a)} Schematic of the problem setup. Inner and outer cylinders at radii $R_i, R_o$ rotating independently with angular speeds $\Omega_i,  \Omega_o$ are maintained at constant temperatures $T_i, T_o$, respectively. Cylinders are assumed to be infinite in the axial ($z$) direction. \aghoredit{(b)} In the $\Omega_i$ versus $\Omega_o$ plane, below the Rayleigh line (shown in red, $q = 2$), perturbations are inviscidly stable, see \citet{rayleigh1917dynamics}. The blue curve represents the viscous instability threshold due to \citet{taylor1923viii}. The green and magenta lines in the first quadrant correspond to Keplerian flows ($q=3/2$) and \aghoredit{pure co-}rotation ($q = 0$, \aghoredit{$\mu = 1$}), respectively. Refer to text for a detailed discussion of the $q$-parameter. \aghoredit{The magenta line in the second quadrant marks pure counter-rotation ($\mu = -1$). In this study, we vary rotation rate ratios in the range $\mu \in [-1, 1]$.} }
 \label{fig:setup}
\end{figure*}

Taylor--Couette flow (TCF) refers to the annular flow between two concentric, independently rotating cylinders. In this paper we consider TCF with radial heating [see Fig. \ref{fig:setup}\aghoredit{(a)}]. 
On account of its rich dynamics and experimental realizability, TCF has served as a test case for pattern-formation theory with a history spanning over a century.
Linear stability analysis is often the first step to understanding the parameter space and demarcating stable and unstable regions under various probing conditions, see \citet{chandrasekhar1961hydrodynamic}, \citet{drazin2004hydrodynamic}. 
Linear stability analysis by \citet{rayleigh1917dynamics} formulated the criterion for inviscid instability, now known as the Rayleigh-criterion. 
According to the Rayleigh-criterion, the inviscid flow is stable (unstable) when the square of circulation increases (decreases) monotonically with radial distance. 
In the $(\Omega_i$,$\Omega_o)$ parameter space, with $\Omega_{i}$ and $\Omega_{o}$ representing rotation rates of the inner and outer cylinders respectively, the condition for inviscid instability can be written as $\Omega_{i}/\Omega_{o} \leq \eta^{-2}$, where $\eta = R_{i}/R_{o}$ is the radius ratio with $R_{i},R_{o}$ being the inner and outer radii, respectively. 
In terms of the rotation rate ratio $\mu = \Omega_o/\Omega_i$, the Rayleigh line that marks the boundary for inviscid instability can be expressed as $\mu = \eta^{2}$ [see Fig. \ref{fig:setup}\aghoredit{(b)}]. 

\citet{taylor1923viii} performed viscous linear stability analysis and found the boundary for the onset of instability in a viscous fluid. 
Taylor also performed experiments and showed that experimental observation was in accordance with the linear stability analysis. 
The first bifurcation from the basic circular Couette flow (CCF) was found to be axisymmetric as the rotation rate of the inner cylinder was increased. 
The axisymmetric toroidal flow that bifurcates from CCF are now known as Taylor vortex flows. 
After the seminal work of \citet{taylor1923viii}, the problem and its variants have remained of interest even today due to experimental viability and a wide range of relevance from industrial applications to geophysical and astrophysical fluid dynamics \cite{gollub1975onset,aghor2021nonlinear,crowley2022turbulence,ji2023taylor}. 
The Rayleigh-criterion is of particular importance in astrophysical flows such as accretion disks. 
The classical Rayleigh-criterion can be violated when other effects such as magnetic fields  \cite{ji2023taylor,nordsiek2015azimuthal} or density stratification \cite{balbus1991powerful, shalybkov2005stability,le2007experimental,le2010strato,robins2020viscous,grossmann2016high} are considered.

The focus of this study is the radially--heated TCF which is important in many problems ranging from rotating machinery in industrial applications \cite{lee1989heat} to geophysical and astrophysical flows \cite{busse1994convection,lopez2013boussinesq,jiang2020supergravitational}.
Early efforts of studying radially heated TCF such as \cite{roesner1978hydrodynamic,soundalgekar1981effects,takhar1985effects} considered axisymmetric disturbances. 
\citet{ali1990stability} studied radially heated Taylor--Couette setup with stationary outer cylinder with axisymmetric \aghoredit{toroidal} as well as non-axisymmetric \aghoredit{helical} disturbances.
Other studies such as \citet{kedia1998numerical},  \citet{meyer2015effect}, \citet{yoshikawa2013instability}, \citet{kang2015thermal,kang2017radial,kang2019numerical} also considered centrifugal buoyancy which was ignored in earlier works. \citet{guillerm2015flow} examined the effect of radial buoyancy in absence of gravitation. \citet{kuo1997taylor} studied the radially heated TCF where they considered finite aspect ratios and studied the effect of varying Grashof number on the full nonlinear system via direct numerical simulations. 
Most of the investigations of radially heated TCF so far have considered the case of rotating inner cylinder and stationary outer cylinder with a few exceptions such as \citet{meyer2021stability} who investigated the effects of centrifugal buoyancy under the assumption of micro-gravity conditions in the Rayleigh-stable regime. 

In this paper we investigate the answers to the following questions: How does outer-cylinder rotation ($\mu \neq 0$) affect the onset of instability? Can the Rayleigh-criterion be violated due to the presence of natural convection? What are the effects of changing the radius--ratio ($\eta$)?

We consider a Boussinesq fluid of density $\rho$, dynamic viscosity $\nu$, thermal diffusivity $\kappa$, and thermal expansion coefficient $\beta$ in a cylindrical annulus of infinite length. 
The acceleration due to gravity ($g$) points in the negative axial direction. 
The Prandtl number ($Pr$) for this study, defined as the ratio of momentum diffusivity and thermal diffusivity, is fixed at unity unless stated otherwise. 
A schematic of the setup is depicted in Fig. \ref{fig:setup}\aghoredit{(a)}. 
The inner cylinder of radius $R_{i}$ spinning with an angular velocity $\Omega_{i}$ is maintained at a temperature $T_{i}$, whereas the outer cylinder of radius $R_{o}$ spinning with an angular velocity $\Omega_{o}$ is maintained at a temperature $T_{o}$. 
\aghoredit{The centrifugal buoyancy term starts becoming important beyond inner Reynolds numbers $Re_i > 5 \times 10^{5}$, see \citet{lopez2013boussinesq}. However, the critical Reynolds numbers (connected to critical Taylor numbers via Eqn. \ref{eq:Ta_Re_relation}) are far lower than this bound. We therefore do not consider the centrifugal buoyancy term in this study. Neglecting the centrifugal buoyancy term is equivalent to working in the limit of vanishing Froude number, when the Froude number is defined as $Fr = \nu/\sqrt{gd^{3}}$ where $d = R_{o} - R_{i}$ is the gap width \cite{yoshikawa2013instability}. 
Furthermore, this also allows us to isolate the effect of outer cylinder rotation.}
We thus arrive at a system similar to \citet{ali1990stability}, but with a general basic azimuthal velocity profile (corresponding to CCF) that incorporates the effect of outer--cylinder rotation.

A note on terminology used in this article follows. We consider $\mu \in [-1, 1]$ ranging from the pure counter--rotation case ($\mu = -1$) to the pure co--rotation ($\mu = 1$) for radius ratios $\eta=0.6$ (wide--gap case) and $\eta=0.9$ (thin--gap case). 
For the rest of this article, we refer to $\mu < 0$ ($\mu > 0$) collectively as the `counter--rotation regime' (`co--rotation regime') and reserve the term `pure counter--rotation case' (`pure co--rotation case') for $\mu = -1$ ($\mu = 1$). In the co--rotation regime, we use the $q$-parameter \cite{ji2006hydrodynamic,nordsiek2015azimuthal} to demarcate regimes in the $(\Omega_i,\Omega_o)$ phase plane. 
The $q$-parameter relates the rotation rate ratio to the radius ratio as
\begin{equation}
    \mu = \frac{\Omega_o}{\Omega_i} = \eta^q.
\end{equation}
Different lines corresponding to different values of $q$ are shown in Fig. \ref{fig:setup}\aghoredit{(b)}. 
The Rayleigh line corresponds to $q = 2$, whereas $q = +\infty, -\infty$ correspond to pure inner and outer cylinder rotations, respectively. 
The Rayleigh-stable regime can be given by $q < 2$. The Rayleigh-stable regime consists of sub-rotation ($\Omega_i < \Omega_o$ or $\mu > 1$, not considered here), super-rotation ($\Omega_i > \Omega_o$ or $\mu < 1$, considered here) and the solid-body rotation line (pure co--rotation case corresponding to $q = 0$). 
The super-rotation regime is also known as quasi-Keplerian regime, since it contains $q = 3/2$ corresponding to Keplerian flows. The quasi-Keplerian regime is thus important for astrophysical flows. 
The Rayleigh line for the wide--gap case is given by $\mu = \eta^{2} = 0.36$, whereas it lies at $\mu = \eta^{2} = 0.81$ for the thin--gap case. 
The Grashof number $\Gr$ is fixed at a value of $1000$ for this study.
For each of the radius ratios, intermediate values of $\mu \in [-1, 1]$ are analyzed.

The rest of the paper is organized as follows. 
In Sec. \ref{sec:2} we formulate the problem by non-dimensionalizing the governing equations and obtain the steady base flow. We then linearize the system around the base flow and use modal ansatz to reduce to an eigenvalue problem in the radial direction. 
We also discuss the symmetries of the system and validate our methodology by comparing results with published literature.
In Sec. \ref{sec:3} we present the results of our analysis for a range of parameters. The analysis reveals interesting regions in the parameter space, namely closed islands of instability where the flow can restabilize with increasing Taylor number. In Sec. \ref{sec:4}, we study the origin and disappearance of these instability islands. We also demonstrate their sensitivity to changes in external parameter $\mu$ for selected azimuthal modes in both wide-- and thin--gaps. Section \ref{sec:5} discusses implications of our analysis. Finally, in Sec. \ref{sec:6} we present conclusions and discuss potential avenues for further investigations. 

\section{Problem Formulation and Numerical Method}
\label{sec:2}

The governing equations for a Boussinesq fluid that control the dynamics of the system are the Navier-Stokes equations (continuity and momentum equations), and the energy equation.
Following \citet{ali1990stability} the governing equations are rendered dimensionless using gap width $d = R_{o} - R_{i}$ as the characteristic length, velocity of the inner cylinder $R_{i}\Omega_{i}$ as the characteristic azimuthal velocity, $U_0 = g\beta d^{2} \DT /\nu$ as the characteristic radial and vertical velocity, $d^{2}/\nu$ as the time scale, $\rho U_0^{2}$ as the characteristic pressure, and the temperature contrast $|\DT| = |T_{i} - T_{o}|$ as the temperature scale. \aghoredit{Asymmetric scaling in the azimuthal ($v$) and meridional ($u,w$) velocities is chosen following \citet{ali1990stability} for the ease of comparison in the natural convection limit of the problem, where no cylinder rotates and the base flow is entirely driven by natural convection. This asymmetric scaling gives rise to the swirl parameter $S = \Omega_{i}R_{i}/U_0$ that appears in the governing equations below. } 

The dimensionless governing equations thus obtained are
\begin{align}
\frac{1}{r} \pd{(ru)}{r} + S\frac{1}{r} \pd{v}{\phi} + \pd{w}{z} = 0, \label{eq:continuity}
\end{align}
\begin{align}
 \pd{u}{t} + \Gr \left[\udotgrad{u} - S^{2}\frac{v^{2}}{r} \right] = -\Gr\pd{p}{r} \nonumber \\+ \left[\left(\delsq - \frac{1}{r^{2}} \right) u  - S\frac{2}{r^{2}}\pd{v}{\phi}\right], \label{eq:r-mom}  
\end{align}
\begin{align}
 S\left[ \pd{v}{t} + \Gr\left(\udotgrad{v} + \frac{uv}{r} \right) \right]  = -\Gr\frac{1}{r}\pd{p}{\phi} \nonumber \\+ \left[ S\left(\delsq - \frac{1}{r^{2}} \right) v  + \frac{2}{r^{2}}\pd{u}{\phi}\right] , \label{eq:phi-mom} 
\end{align}
\begin{align}
\pd{w}{t} + \Gr \left[\udotgrad{w}  \right] = -\Gr \pd{p}{z} + \delsq w  + T, \label{eq:z-mom} \end{align}
\begin{align}
\pd{T}{t} + \Gr \left[\udotgrad{T}  \right] = \frac{1}{Pr} \delsq T, \label{eq:energy}
\end{align}
where $u, v, w$ are the dimensionless velocities in the radial ($r$), azimuthal ($\phi$), and axial ($z$) directions respectively. Here, $p$ and $T$ are dimensionless pressure and temperature respectively, and $\delsq$ is the Laplacian operator defined as  
\begin{eqnarray}\label{eq:delsq}
\delsq \equiv \pdd {}{r} + \frac{1}{r}\pd{}{r} + \frac{1}{r^{2}}\pdd{}{\phi} + \pdd{}{z}.
\end{eqnarray}
The nondimensionalization gives rise to the dimensionless parameters Prandtl ($Pr$), Taylor ($\Ta$) and  Grashof ($\Gr$) numbers given by
\begin{align}
Pr = \frac{\nu}{\kappa}, \quad \Ta = \frac{2 \eta^{2} \Omega_{i}^{2} d^{4}}{\nu^{2}(1-\eta^{2})}, \quad \Gr = \frac{g \beta \Delta T d^3}{\nu^2}.
\end{align}
The Taylor number $\Ta$ is proportional to the rotation rate of the inner cylinder and can also be related to the Reynolds number based on inner cylinder rotation as
\begin{eqnarray}\label{eq:Ta_Re_relation}
\Ta = \frac{2 (1-\eta)}{(1+\eta)} Re_{i}^{2},
\end{eqnarray}
where $Re_{i} = \Omega_{i} R_{i} d /\nu$ is the inner Reynolds number. 
The swirl parameter $S = \Omega_{i}R_{i}/U_0$ is not an independent parameter. It can be expressed in terms of other dimensionless parameters as
\begin{eqnarray}\label{eq:S}
S = \frac{\left(Ta(1+\eta)/2(1-\eta) \right)^{1/2}}{\Gr} = \frac{Re_{i}}{\Gr}.
\end{eqnarray}
The classical isothermal TCF with all velocities scaled by $\Omega_i R_i$ is recovered by substituting $T = 0$ and $S = 1$, replacing $\Gr$ with $Re_{i}$.
Infinite cylinders are assumed and Dirichlet boundary conditions in the radial direction are imposed, i.e.,
\begin{eqnarray}\label{eq:bc}
 &&(u, v, w, T) = (0, 1, 0, 1) \textrm{ at } r = r_{i} = \frac{\eta}{1-\eta}, \\
 &&(u, v, w, T) = \left(0, \frac{\mu}{\eta}, 0, 0\right) \textrm{ at } r = r_{o} = \frac{1}{1-\eta}, 
 \end{eqnarray}
 with $r_{i} = R_{i}/d$, $r_{o} = R_{o}/d$ being the dimensionless inner and outer radii respectively. In the next section, we formulate the base state and linearize the system in order to perform a linear stability analysis.

\subsection{Base flow and linearization}

The base state can be obtained analytically for this problem assuming radial dependence of the steady basic fields. \aghoredit{Experiments by \citet{eckert1961natural} identified three regimes in both planar and cylindrical gaps, namely conduction, transition and convection. When the flow is in the convection regime, a vertical temperature gradient must also be considered along with the radial temperature gradient. \citet{ali2005linear} have analysed linear stability of radially heated Taylor--Couette flow in the convection regime for the case of stationary outer cylinder. Here, we assume infinite axial extent and the base flow corresponds to the conduction regime with no vertical thermal stratification.}
For the basic azimuthal flow ($\vb$), the well-known general CCF profile in the azimuthal direction valid for nonzero $\mu$ is obtained, see for example, \citet{chandrasekhar1961hydrodynamic}. In the case of $\mu = 0$ corresponding to the stationary outer cylinder, the basic azimuthal velocity profile reduces to the one considered by \citet{ali1990stability}. In the classical isothermal Taylor-Couette setup, there is no axial flow in the base state. However, in the presence of a radial temperature gradient an axial flow is induced due to natural convection, since gravity is perpendicular to the imposed temperature gradient \cite{choi1980stability}. 
As the axial flow $\wb$ is only dependent on the basic temperature profile, outer cylinder rotation ($\mu \neq 0$) does not alter the basic axial flow and is thus the same as reported in Ali and Weidman \cite{ali1990stability}. 
The base flow can be obtained in closed form analytically, given below:
\begin{align}
\ub = 0, \quad \vb = Ar + B/r, \quad 
\Tb = \frac{\ln{\left[(1-\eta) r\right]}}{\ln{\eta}}, \nonumber \\
\wb = \frac{1}{(1-\eta)^{2}} \frac{C}{D} \left[ (1-\eta)^{2} r^{2} - 1 + (1-\eta^{2}) T_{b} \right] \nonumber \\- \frac{1}{4(1-\eta)^{2}} \left[(1-\eta)^{2}r^{2} - \eta^{2}\right]  T_{b}, \nonumber   
\end{align}
with
\begin{align}
 A = \frac{\mu - \eta^{2}}{\eta(1+\eta)}, \quad
 B = \frac{\eta(1-\mu)}{(1-\eta)(1-\eta^{2})}, \quad \nonumber \\
 C = (1-\eta^{2})(1-3\eta^{2}) - 4 \eta^{4}\ln{\eta}, \nonumber \\
 D = 16\left[(1-\eta^{2})^{2} + (1-\eta^{4})\ln{\eta} \right]. \nonumber
\end{align}
We perturb the primitive variables as
\begin{eqnarray} \label{eq:perturb}
    \left[u, v, w, T, p \right] = [\ub, \vb, \wb, \Tb, \pb] (r) + \left[ u', v', w', T', p' \right],
\end{eqnarray}
and substitute into the governing equations \eqref{eq:continuity} -- \eqref{eq:energy}. Thereafter, imposing the boundary conditions and neglecting higher-order terms in the perturbations we obtain the linearized system of governing equations as
\begin{align}
\begin{split}
 \frac{1}{r} \pd{(ru)}{r} + \frac{S}{r} \pd{v}{\phi} + \pd{w}{z}  = 0, \label{eq:lincontinuity}
 \end{split}
\end{align}

\begin{align}
\begin{split}
 \pd{u}{t} + \Gr \left[S \frac{\vb}{r} \pd{u}{\phi} + \wb \pd{u}{z} - S^{2} \frac{2\vb v}{r} \right] = -\Gr\pd{p}{r} \\+ \Lapr, \label{eq:lin-r-mom} 
 \end{split}
\end{align}

\begin{align}
\begin{split}
 S\left[ \pd{v}{t} + \Gr\left( u \frac{d \vb}{dr}+ S \frac{\vb}{r} \pd{v}{\phi} + \wb \pd{v}{z} + \frac{u \vb}{r} \right) \right] = -\Gr\frac{1}{r}\pd{p}{\phi} \\+ \Lapphi, \label{eq:lin-phi-mom} \\ 
 \end{split}
\end{align}

\begin{align}
\begin{split}
 \pd{w}{t} + \Gr \left[u \frac{d\wb}{dr} + S \frac{\vb}{r} \pd{w}{\phi} + \wb \pd{w}{z}  \right] = -\Gr \pd{p}{z} + \delsq w  + T, \label{eq:lin-z-mom} \\
 \end{split}
\end{align}

\begin{align}
\begin{split}
 \pd{T}{t} + \Gr \left[u\frac{d\Tb}{dr} + S \frac{\vb}{r} \pd{T}{\phi} + \wb \pd{T}{z} \right] = \frac{1}{Pr} \delsq T.\label{eq:lin-energy}
 \end{split}
\end{align}
Note that we have dropped primes on the perturbation fields for brevity. From here onward, $[u, v, w, p, T]$ refer to the perturbation fields. 

The boundary conditions become
\begin{eqnarray}
 &(u, v, w, T) = (0, 0, 0, 0) \textrm{ at } r = r_{i}, r_{o}.
 \label{eq:linbc}
 \end{eqnarray}
Substituting the modal perturbation ansatz 
\begin{align}\label{eq:modal_ansatz}
  [u, v, w, T, p] = [\hat{u}, \hat{v}, \hat{w}, \hat{T}, \hat{p}]  (r) \, \exp{\left[i(k_z z + m \phi) + \sigma t\right]} + c.c.,
\end{align}
in Eqs. \eqref{eq:lincontinuity}-\eqref{eq:linbc}, a generalized eigenvalue problem of the form $\bsym{A} \bx = \sigma \bsym{B} \bx$ in $r$ is obtained, with $\sigma$ as the eigenvalue and $\bx = [\hat{u}, \hat{v}, \hat{w}, \hat{T}, \hat{p}]  (r)$. Here `c.c.' stands for `complex conjugate'. Since the domain in the azimuthal direction has a natural periodicity of $2\pi$, the azimuthal wavenumber $m$ only takes integer values. On the other hand, since the cylinders are assumed to be infinite in the axial direction, the axial perturbation wavenumber $k_z$ is a continuous parameter and can take real values. The eigenvalue $\sigma$ is in general a complex number of the form $\sigma = \sigma_{r} + i \sigma_{i}$, with $\sigma_{r}, \sigma_{i} \in \mathcal{R}$. The real part of the eigenvalue $\sigma_{r}$ represents the growth rate and the imaginary part $\sigma_{i}$ represents the frequency of the evolution of perturbations according to Eq. \eqref{eq:modal_ansatz}. 
Linearly stable perturbations are characterized by a negative growth rate ($\sigma_r < 0$), whereas linearly unstable perturbations have a positive growth rate ($\sigma_r > 0$). The locus of points satisfying $\sigma_r = 0$ marks the neutral stability boundary which denotes the transition between stable and unstable regimes, see for example \citet{drazin2004hydrodynamic}. 
\aghoredit{Critical perturbation modes correspond to the minima of the neutral stability curves. We denote the eigenvalues corresponding to the critical perturbation modes with a subscript `c', i.e., $\sigma_c = \sigma_{rc} + i \sigma_{ic}$. By definition, $\sigma_{rc} = 0$ due to zero growth rate and $\sigma_c = i \sigma_{ic}$.}
As mentioned in \citet{ali1990stability}, critical perturbation modes can be completely described by the triplet $(m, k, \sigma_{ic})$. The non-dimensional axial propagation speed of the phase lines $C$, wavelength $\lambda$ normal to lines of constant phase, and inclination of the phase lines $\psi$ with respect to the horizontal are given by
\begin{align}\label{eq:perturbation_shape}
    C = -\frac{\sigma_{ic}}{k_z}, \quad \lambda = \frac{2\pi}{\left( m^{2}/r^{2} + k_z^{2} \right)^{1/2}}, \quad \psi = - \tan^{-1} \left( {\frac{m}{rk_z}} \right).
\end{align}
If $\sigma_{ic} = 0$, the critical modes are stationary in time, whereas if $\sigma_{ic} \neq 0$, the critical modes are oscillatory in time according to Eqn. \ref{eq:modal_ansatz}.

Note that as the linearized equations \eqref{eq:lin-r-mom} -- \eqref{eq:lin-energy} each have a second derivative term with respect to $r$, the resulting eigenvalue problem is $8$-dimensional \aghoredit{under the first-order reduction } and is closed by $8$ boundary conditions given in Eqs. \eqref{eq:linhatbc}. \aghoredit{After discretization in the radial direction with $N$ grid-points, the numerical eigenvalue problem becomes $8N$ dimensional.}
In this study, we solve the resulting eigenvalue problem using `eigentools' package from Dedalus \cite{burns2020dedalus}. 
Eigentools is equipped with automatic rejection of spurious modes by calculating the drift ratio \cite{boyd2001chebyshev}. 
This is performed by comparing the eigenvalues at a specified resolution and a higher resolution of 1.5 times the original resolution, see \citet{oishi2021eigentools} for details. We use Chebyshev polynomials to discretize in the radial direction. 
Substituting the modal ansatz Eq. \eqref{eq:modal_ansatz} into Eq. \eqref{eq:linbc} appropriate homogeneous Dirichlet boundary conditions are obtained for the eigevnalue problem 
\begin{eqnarray} \label{eq:linhatbc}
(\hat{u}, \hat{v}, \hat{w}, \hat{T})  = (0, 0, 0, 0) \textrm{ at } r = r_{i}, r_{o}.   
\end{eqnarray}
The calculations in this study are performed with 48 grid-points in the radial direction. To ensure convergence, we ran some calculations again at a higher grid--size with 64 grid--points and no significant difference was observed. \aghoredit{ In the wide gap, we re--ran the calculations of neutral curves for $m = \pm 2$ at $\mu = -1$ and $m = -4$ for $\mu = 1$ with 64 grid--points in the radial direction. In the thin gap, we chose $ m = 2$ at $\mu = -1$ and $m = -20$ at $\mu = 1$ cases to re--run at 64 grid--points in the radial direction. The values are chosen for critical cases corresponding at the extremities of our range of rotation rate ratio. The value for $m = -2$ at $\mu = -1$ in the wide gap was chosen since it shows a neutral stability curve with two minima, ensuring that 48 grid points are enough to capture non trivial aspects of neutral curves.}
\subsection{Symmetries}
We briefly discuss symmetries of the linearized system [Eqs. \eqref{eq:lincontinuity}--\eqref{eq:lin-energy}] and their consequences, and justify our choice of numerical parameters in this section. 
Two symmetries of the linearized perturbation equations for the stationary outer cylinder ($\mu = 0$) were identified previously \cite{ali1990stability}. 
We generalize these symmetries to include any nonzero $\mu$ as 
\begin{align}
  S_1 = S^{c/r} (\Omega): [\hat{u}, \hat{v}, \hat{w}, \hat{T}, \hat{p}; Ta, \Gr, k_z, \sigma, \mu, m] \rightarrow \nonumber\\ [\hat{u}, -\hat{v}, \hat{w}, \hat{T}, \hat{p}; Ta, \Gr, k_z, \sigma, \mu, -m], \label{eq:symms1} 
\end{align}

\begin{align}
  S_2 = S^{c/r} (\DT): [\hat{u}, \hat{v}, \hat{w}, \hat{T}, \hat{p}; Ta, \Gr, k_z, \sigma, \mu, m] \rightarrow \nonumber\\ [-\hat{u}^{*}, \hat{v}^{*}, \hat{w}^{*}, \hat{T}^{*}, \hat{p}^{*}; Ta, -\Gr, k_z, \sigma^{*}, \mu, -m]. \label{eq:symms2}
\end{align}
\aghoredit{
In the notation of \citet{ali1990stability}, the superscript ${c/r}$ in the labels for symmetries stands for the (naturally) "convecting/rotating" system.}
The symmetry $S_1 = S^{c/r} (\Omega)$ corresponds to a situation where the inner cylinder is rotated in the opposite direction, i.e., $\Omega_i \rightarrow -\Omega_i$.
For the general case considered in our study with $\mu$ allowed to be nonzero, if we consider $\Omega_o \rightarrow -\Omega_o$, keeping $\mu$ invariant, the modified version of $S^{c/r} (\Omega)$ given in Eq. \eqref{eq:symms1} can be shown to hold. 
It is seen that only $\vb \rightarrow -\vb$ is sufficient to quantify the effect of this transformation on the base state whereas boundary conditions for the perturbation fields remain the same as Eq. \eqref{eq:linbc}. 

The symmetry $S_2 = S^{c/r} (\DT)$ represents the scenario with temperature gradient reversed, i.e., $\Gr \rightarrow -\Gr$. 
We first note that $\Gr \rightarrow -\Gr$ does not alter the basic temperature profile $\Tb$ as it is nondimensionalized using the temperature contrast $|\DT| = |T_{i} - T_{o}|$. 
As a consequence, $\wb$ (which is completely determined by $\Tb$) is also invariant under $\Gr \rightarrow -\Gr$. 
Substituting $S^{c/r} (\DT) [\hat{u}, \hat{v}, \hat{w}, \hat{T}, \hat{p}; Ta, \Gr, k_z, \sigma, \mu, m]$ in Eq. \eqref{eq:modal_ansatz} and Eqs. \eqref{eq:lincontinuity}-\eqref{eq:lin-energy}, we obtain the complex conjugated version of the original system. 
This is independent of whether $\mu$ is zero or otherwise. 
Hence the $S^{c/r} (\DT)$ symmetry can be extended for nonzero $\mu$ as well, implying that for each $\Gr > 0$ there exists an equivalent case of $\Gr < 0$ with the inner wall being cooler than the outer wall.

Using a combination of these two symmetries, a third symmetry is identified as $S_3 = S_1 \cdot S_2 = S_2 \cdot S_1$ where
\begin{align}\label{eq:S3}
    S_3: [\hat{u}, \hat{v}, \hat{w}, \hat{T}, \hat{p}; Ta, \Gr, k_z, \sigma, \mu, m] \rightarrow \nonumber \\ [-\hat{u}^{*}, -\hat{v}^{*}, \hat{w}^{*}, \hat{T}^{*}, \hat{p}^{*}; Ta, -\Gr, k_z, \sigma^{*}, \mu, m].
\end{align}
The effect of these symmetries can be explained as follows. For each solution of the linearized equations with $(\Omega_i, \DT, \mu)$, there are 3 equivalent solutions for the cases $(-\Omega_{i}, \DT, \mu)$, $(\Omega_{i}, -\DT, \mu)$ and $(-\Omega_{i}, -\DT, \mu)$. If the first case corresponds to spirals having phase speed and inclination with respect to the horizontal given by $(C, \psi)$, then the $S_1, S_2, S_3$ symmetries imply that there also exist solutions with $(C, -\psi)$, $(-C, -\psi)$ and $(-C, \psi)$. We have therefore generalized the symmetries identified by \citet{ali1990stability} to include nonzero outer-cylinder rotation. 
We only consider $\Gr > 0$ in our analysis. Other equivalent cases can be constructed for $\Gr < 0$ using symmetry arguments. 
\subsection{Validation} \label{sec:validation}
\begin{figure}
 \begin{center}
  \includegraphics[scale=0.5]{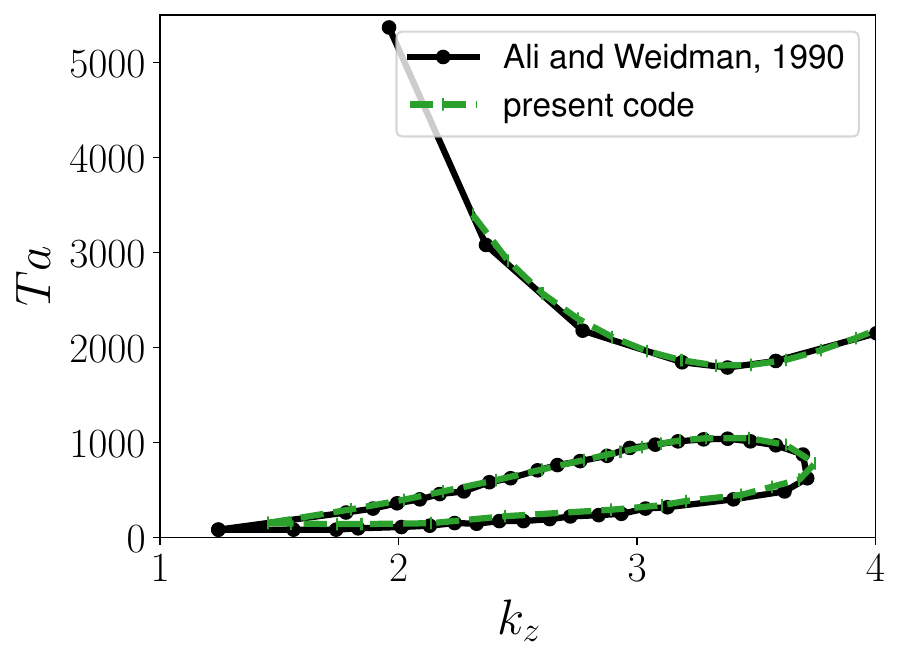}
\end{center}
\caption{A comparison of the present code with results from published literature at $Pr = 15, \eta = 0.6, \Gr = 300, m = -2$ in our notation. Dashed green lines with vertical markers represent neutral stability data obtained from the current methodology and black circles represent data extracted from \citet{ali1990stability}.}
 \label{fig:compare}
\end{figure}

As mentioned earlier, we use eigentools package from Dedalus to solve the resulting eigenvalue problem. 
To verify our linear stability routine, we compare neutral stability curves obtained from our code to those reported in \citet{ali1990stability} for the case of stationary outer cylinder ($\mu = 0$). 
A sample calculation for $m = -2$ mode is shown in Fig. \ref{fig:compare} at $Pr=15, \Gr=300$ and $\eta=0.6$. It can be seen that the linear stability routine accurately predicts both open and close neutral stability boundaries. 

\section{Linear Stability Analysis}
\label{sec:3}

 \begin{figure*}
  \includegraphics[width = 0.98\textwidth]  {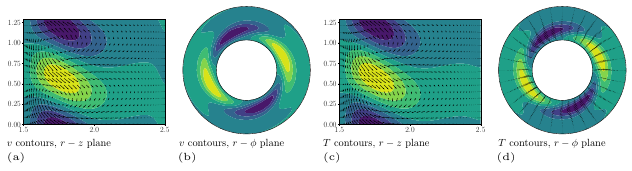}
\caption{
Normalized critical eigenfunctions for $\eta=0.6, \Gr = 1000$ for counter--rotation case with $\mu = -1, m=2$; slices of contour plots for the critical azimuthal velocity eigenmode in the \aghoredit{(a)} $r - z$ and \aghoredit{(b)} $r - \phi$ planes, followed by contour plots of temperature eigenmode in the \aghoredit{(c)} $r - z$ and \aghoredit{(d)} $r - \phi$ planes. The quiver plots in the $r - z$ slices are made up of corresponding eigenmodes for meridional velocity perturbations ($u$ and $w$).}
 \label{fig:most_unstable_eta_0.6_G_1000_counterrot}
\end{figure*}
\begin{figure*}
 \includegraphics[width = 0.98\textwidth]  {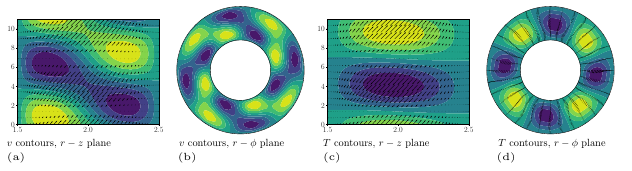}
\caption{
Normalized critical eigenfunctions for $\eta=0.6, \Gr = 1000$ for co--rotation case with $\mu = 1, m=-4$; slices of contour plots for the critical azimuthal velocity eigenmode in the \aghoredit{(a)} $r - z$ and \aghoredit{(b)} $r - \phi$ planes, followed by contour plots of temperature eigenmode in the \aghoredit{(c)} $r - z$ and \aghoredit{(d)} $r - \phi$ planes. The quiver plots in the $r - z$ slices are made up of corresponding eigenmodes for meridional velocity perturbations ($u$ and $w$).}
 \label{fig:most_unstable_eta_0.6_G_1000_corot}
\end{figure*}

In this section, we present the results of the linear stability analysis. 
As mentioned earlier, we analyze the parameters $Pr=1, \Gr = 1000$ and radius ratios $\eta=\{0.6, 0.9 \}$ which correspond to wide and thin--gap cases, respectively. 
We sweep over the rotation rate ratio ($\mu$) and azimuthal wave number ($m$) to locate the neutral stability boundaries between stable and unstable regimes. 
For a given combination of control parameters (radius ratio, the rotation rate ratio and azimuthal perturbation wavenumber), the Taylor number corresponding to the onset of instability is recorded in Table \ref{tbl:Gr_1000_neutral}.
The table also records the fastest growing mode for a given combination of control parameters, shown by highlighted entries. 
We now discuss critical modes at the onset of instability for wide-- and thin--gaps. 
For both cases, closed regions of instability are found. 
We refer to these closed regions of instability as instability islands or closed disconnected neutral curves (CDNCs) and also discuss their possible origin, disappearance and/morphing into open neutral stability curves. 

\subsection{Wide--gap $\eta = 0.6$}
For the wide--gap annulus i.e., $\eta=0.6$, the critical Taylor numbers $\Ta_c$ and corresponding critical axial wave numbers ($\kc$) are reported in Table \ref{tbl:Gr_1000_neutral}.
For each $\mu \in \{-1,-0.5,-0.2,0,0.2,0.5,1\}$ we calculate the critical Taylor number $\Ta_c$ and the critical frequency $\sigma_{ic}$ for different azimuthal modes.
We find the fastest growing mode as the one with the smallest $\Ta_c$.
These values are highlighted in the Table \ref{tbl:Gr_1000_neutral}.
In all the cases reported the modes at the onset of instability were found to be oscillatory, evident from nonzero critical frequency $\sigma_{ic}$. 
Figures \ref{fig:most_unstable_eta_0.6_G_1000_counterrot} and \ref{fig:most_unstable_eta_0.6_G_1000_corot} depict the fastest growing modes for the pure counter--rotation ($\mu = -1$) and pure co--rotation ($\mu = 1$) cases, corresponding to extreme values of $\mu$ considered here. 

For the pure counter--rotation case (Fig. \ref{fig:most_unstable_eta_0.6_G_1000_counterrot}) the temperature and azimuthal velocity eigenmodes can be seen to be confined near the inner cylinder. 
It should be noted that this behavior for the pure counter--rotation case is also typical of lower values of $\Gr$ such as $100,500$ (not reported here).
For the pure co--rotation case, although the temperature and velocity modes reach all the way to the outer cylinder (Fig. \ref{fig:most_unstable_eta_0.6_G_1000_corot}), an interesting \aghoredit{``}bimodal radial structure\aghoredit{''} of the azimuthal velocity eigenmode is observed. \aghoredit{This bimodal radial structure of the azimuthal velocity eigenmode has two maxima in the radial coordinate, as opposed to the one observed in the case of pure counter--rotation. This  gives rise to the azimuthal velocity contours as shown in Fig. \ref{fig:most_unstable_eta_0.6_G_1000_corot} (a). These contours of azimuthal velocity eigenmode are qualitatively different than ones obtained with only one maximum in the radial direction, for example, the ones shown in Fig. \ref{fig:most_unstable_eta_0.6_G_1000_counterrot}(a).} 

\begin{figure*}
 \begin{center}
  \includegraphics[width=0.98\textwidth]{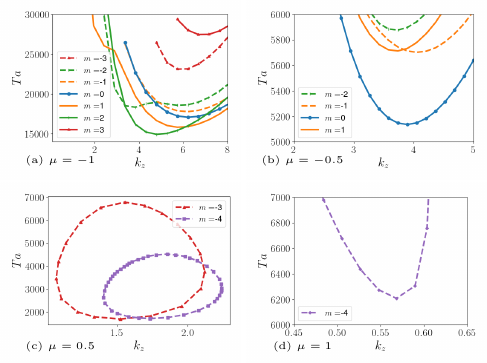}
\end{center}
\caption{Neutral stability curves for $\eta=0.6, \Gr = 1000$ for different azimuthal perturbation numbers with $\mu = $ 
\aghoredit{(a)} $-1$, \aghoredit{(b)} $-0.5$, \aghoredit{(c)} $0.5$ and \aghoredit{(d)} $1$. For open neutral stability curves unstable regions lie above the neutral stability boundary. The closed neutral stability curves enclose islands of instability. Regions outside the closed boundaries are stable. For quantitative details, 
see Table \ref{tbl:Gr_1000_neutral}.}
 \label{fig:neutral_stab_G_1000_eta_0.6}
\end{figure*}

Figure \ref{fig:neutral_stab_G_1000_eta_0.6} plots neutral stability curves for four selected values of the rotation rate ratio. Here, except for the case of $\mu =-0.5$ in the counter--rotation regime, the fastest-growing modes are non-axisymmetric. \aghoredit{Neutral stability curves for show a non--monotonic behavior with respect to the azimuthal wavenumber $m$. For example, for pure counter--rotation ($\mu = -1$), critical Taylor numbers for azimuthal modes $m = 0, 1, 2$ decrease, but the critical Taylor number for $m = 3$ is seen to be more than that of the axisymmetric mode. Furthermore, for $m = -2$, we observe two minima. Both these behaviors might have their explanation in the existence of two distinct open neutral curves at some other values of parameters which have merged into one another at the parameter values considered in this study. A more detailed investigation might show the process of merging of two open neutral curves, much like merging of open and closed neutral stability curves tracked in Figs. \ref{fig:island_eta_0.6_m_-3} and \ref{fig:island_origin_of_small_k_instab}.}

\aghoredit{Positive and negative azimuthal modes can be seen to have different linear stability thresholds from Fig. \ref{fig:neutral_stab_G_1000_eta_0.6}. This behavior can be explained from a symmetry breaking perspective, see Sec. \ref{sec:symm_breaking} for more discussion.}
Interestingly, we find unstable modes even in the Rayleigh-stable regime $\mu > \eta^{2} = 0.36$. 
For $\mu = 0.5$, CDNCs are found, see Fig. \ref{fig:neutral_stab_G_1000_eta_0.6} \aghoredit{(c)}. The flow is unstable inside the CDNCs and stable outside. 
However, for the pure co--rotation case ($\mu = 1$) we do not observe CDNCs, nor could we observe any unstable region for $m = -3$. Instead, we find $m = -4$ to be the fastest-growing mode with a seemingly open neutral stability boundary. The origin and disappearance of these instability islands is further discussed in Sec. \ref{sec:4}.

\subsection{Thin--gap, $\eta = 0.9$}
We now discuss the thin--gap annulus of $\eta = 0.9$. 
Table \ref{tbl:Gr_1000_neutral} summarizes the results of the linear stability analysis.
As seen from Table \ref{tbl:Gr_1000_neutral}, one often needs to go beyond $m = -4$ to find the fastest growing mode for many rotation rate ratios. 
Here, the Rayleigh-line is given by $\mu = \eta^{2} = 0.81$ and the Rayleigh stable regime is given by $\mu > \eta^{2} = 0.81$. 
Similar to the wide--gap case, all the modes at the onset of instability are found to be oscillatory, evident from nonzero critical frequencies $\sigma_{ic}$ in Table \ref{tbl:Gr_1000_neutral}.

\begin{figure*}
  \includegraphics[width = 0.98\textwidth]  {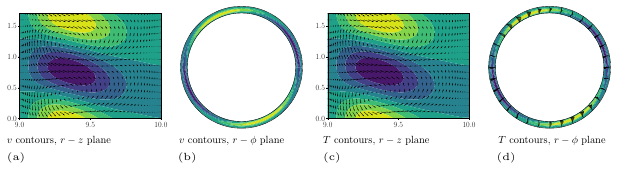}
\caption{
Normalized critical eigenfunctions for $\eta=0.9, \Gr = 1000$ for counter--rotation case with $\mu = -1, m=2$ --  colors and slices to be interpreted as in Fig. \ref{fig:most_unstable_eta_0.6_G_1000_counterrot}.}
\label{fig:most_unstable_eta_0.9_Gr_1000_counterrot}
\end{figure*}

\begin{figure*}
  \includegraphics[width = 0.98\textwidth]  {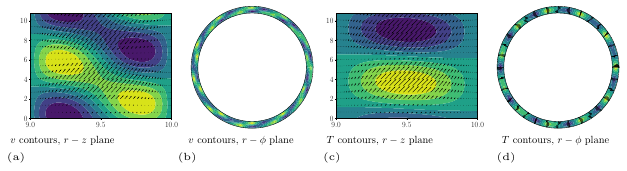}
\caption{
Normalized critical eigenfunctions for $\eta=0.9, \Gr = 1000$ for co--rotation case with $\mu = 1, m=-20$ --  colors and slices to be interpreted as in Fig. \ref{fig:most_unstable_eta_0.6_G_1000_counterrot}.}
 \label{fig:most_unstable_eta_0.9_G_1000_corot}
\end{figure*}

As in the wide--gap case, we plot eigenfunctions at $\mu=-1$ (pure counter--rotation) and $\mu = 1$ (pure co--rotation). 
 \begin{figure*}
 \begin{center}
  \includegraphics[width=0.98\textwidth]{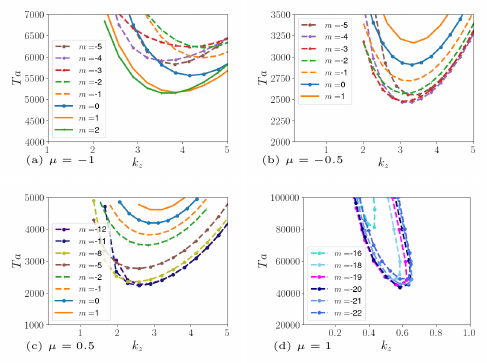}
\end{center}
\caption{Neutral stability curves for $\eta=0.9, \Gr = 1000$ for different azimuthal perturbation numbers with $\mu = $ 
\aghoredit{(a)} $-1$, \aghoredit{(b)} $-0.5$, \aghoredit{(c)} $0.5$, \aghoredit{(d)} $1.0$.
From the neutral stability curves at $\Gr = 1000$ for $\mu = \{-1, -0.5, 0.5, 1\}$.}
 \label{fig:neutral_stab_G_1000_eta_0.9}
\end{figure*}
Figures \ref{fig:most_unstable_eta_0.9_Gr_1000_counterrot} and \ref{fig:most_unstable_eta_0.9_G_1000_corot} plot the fastest growing marginal modes at $\Gr=1000$ with $m = 2, -20$ for $\mu=-1, 1$, respectively.  
Similar to the wide--gap case, here too, the fastest growing modes for $\mu = -1$ are confined near the inner cylinder in the radial direction due to counter--rotation. 
\aghoredit{For the pure co--rotation case, the fastest growing mode is observed to be $m = -20$. The azimuthal velocity eigenmode has a similar radial variation as that of $m = -4$ in the wide gap. In particular, instead of having one maximum in the radial coordinate, it has two maxima. On the other hand, the temperature eigenfunction has only one maximum in the radial coordinate. The effect of this ``bimodal'' variation in the radial coordinate of the azimuthal velocity eigenmode is shown in Fig. \ref{fig:most_unstable_eta_0.9_G_1000_corot}(a), where the contours show two maxima in the radial direction, instead of one seen in the pure counter--rotation case, see Fig. \ref{fig:most_unstable_eta_0.9_Gr_1000_counterrot}(a).}
The critical values of Taylor number found in the thin--gap case are, in general, lower than their wide--gap counterparts at the same value of rotation rate ratio, see Table \ref{tbl:Gr_1000_neutral} for comparison. For the values of rotation rate ratios shown in Fig. \ref{fig:neutral_stab_G_1000_eta_0.9}, no CDNCs were located in our parameter sweep. However, we found CDNCs for many intermediate values of $\mu$ for the critical mode in the pure co--rotation case $m = -20$. These CDNCs are shown in Fig. \ref{fig:island_origin_of_small_k_instab} \aghoredit{(b)} and are further discussed in Sec. \ref{sec:4}.

\section{Variation of instability islands with the rotation rate ratio} \label{sec:4}
Instability islands/CDNCs have been found for a variety of physical problems such as natural convection \cite{chen1989stability}, multicomponent convection \cite{pearlstein1989onset,lopez1990effect,shankar2022stability}, laterally heated cylindrical convection \cite{wang2014linear}, annular Poiseuille flow \cite{cotrell2004connection,cotrell2006linear}, and strato--rotational instability \cite{robins2020viscous}, to name a few.   
From these investigations, it is clear that CDNCs are ubiquitous and their effects on the critical parameters can be dramatic.

The flow is unstable inside the closed regions and stable outside. 
Unlike the open neutral stability curves these instability islands are peculiar, as they guide the regions in the parameter space where increasing Taylor number can restabilize the flow. 
For TCF, \citet{ali1990stability} found instability islands in the case of pure inner cylinder rotation for the radially heated TCF, for example, see Fig. \ref{fig:compare}. 
We also find islands of instability for some rotation rate ratios and azimuthal perturbation wave-numbers for both wide--gap and thin--gap cases. 

In this section, we investigate the origin, disappearance, or morphing of these islands in the parameter space as the rotation rate ratio $\mu$ is varied while radius ratio is held fixed. 
We must note that it is difficult to obtain instability islands in the parameter space, since they can be too small and might not be covered in the parameter sweep. 
Thus, finding all CDNCs is an extensive task.
Therefore, we focus on two important modes in the wide--gap case, namely $m = {-3, -4}$ (the only modes that we find with instability islands in the Rayleigh--stable regime) and the mode $m = -20$ (the critical mode in the pure co--rotation case) in the thin--gap case. 
These cases serve as examples of how these regions can change as the rotation rate ratio $\mu$ is varied and sufficiently demonstrate the sensitivity of the instability islands to the control parameter $\mu$.  
\begin{figure*}
 \begin{center}
  \includegraphics[width = 0.98\textwidth]{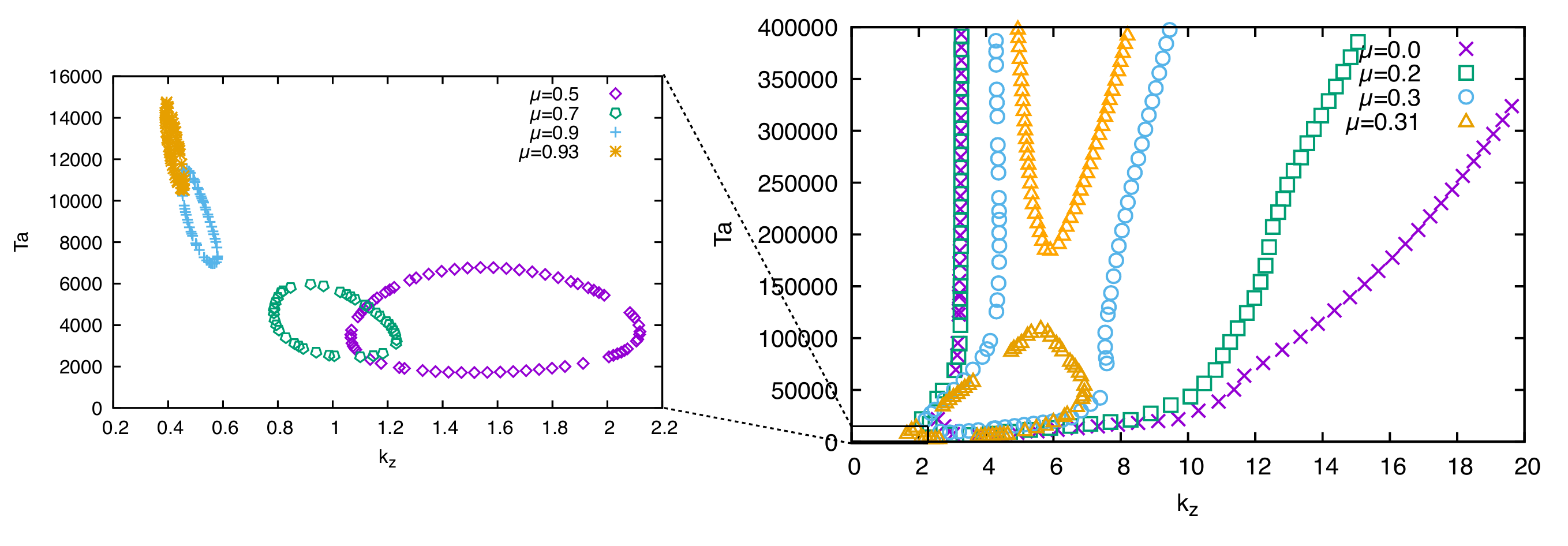}
\end{center}
\caption{Neutral stability curves at different values of $\mu$ for $m = -3$ mode in the wide--gap case. We did not find open neutral stability curves for the values of $\mu$ in the \aghoredit{left inset} figure up to $Ta = 4 \times 10^5$ and $k_z = 20$.
An instability island separates from the open neutral stability curves at $\mu \approx 0.305$ and is seen to move in towards small-$k_z$ with increasing $\mu$. The island disappears at $\mu \approx 0.95$.}
 \label{fig:island_eta_0.6_m_-3}
\end{figure*}
\begin{figure*}
 \begin{center}
  \includegraphics[width = 0.98\textwidth]{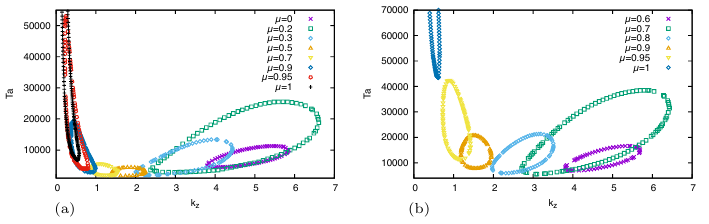}
\end{center}
\caption{Neutral stability curves at different values of $\mu$ for \aghoredit{(a)} $m = -4$ mode in the wide--gap case and \aghoredit{(b)} $m = -20$ in the thin--gap case. The (seemingly) open neutral stability boundary corresponding to an onset of long--wavelength instability in the pure co--rotation case ($\mu = 1$) can be traced back to the closed islands of instability as $\mu$ is decreased below $1$.}
 \label{fig:island_origin_of_small_k_instab}
\end{figure*}

Figures \ref{fig:island_eta_0.6_m_-3} plots the neutral stability curves for mode $m = -3$ in the wide--gap as $\mu$ is varied. 
It is seen from the figure that the left and right boundaries of the (seemingly) open neutral stability curve at $\mu = 0$ come closer as $\mu$ is increased. 
At $\mu \approx 0.305$ an instability island emerges and at $\mu = 0.31$ we see two distinct regimes containing a CDNC and an open neutral stability curve. When both CDNCs and semi-infinite open neutral stability boundary exist, three critical Taylor numbers should be studied via DNS -- (i) the lowest Taylor number of the CDNC, (ii) the one corresponding to re--stabilization and (iii) the one corresponding to onset of instability for the open neutral stability curve. However, we only report Taylor numbers corresponding to case (i) in Table \ref{tbl:Gr_1000_neutral}. An investigation into the splitting process of the two types of neutral stability boundaries is out of scope of the present paper and is left as future work. 
As $\mu$ is varied from $0.31$ through $0.5$, the island shrinks in size and then starts stretching in the direction of the Taylor-number axis. 
At the same time, it seems to occupy an increasingly thinner band of axial perturbation wavenumbers $k_z$.
For example, the island at $\mu = 0.7$ lies between $2000 < Ta < 6000$ and $ 0.7 < k_z < 1.3$, whereas the island at $\mu = 0.9$ exists for $6000 < Ta < 12000$ and $ 0.4 < k_z < 0.6$. 
At $\mu = 0.93$, the island has an even thinner extent in the axial wavenumbers, $0.4 < k_z < 0.47$ and disappears at $\mu \approx 0.95$. This explains why we could not locate any unstable region for the $m=-3$ mode for the pure co--rotation case in the wide--gap.

Figure \ref{fig:island_origin_of_small_k_instab} \aghoredit{(a)} plots neutral stability curves for $m = -4$ mode for different rotation rate ratios. 
\aghoredit{Here, an island is at $\mu = 0$ first increases in size when the rotation rate ratio is increased to $\mu = 0.2$ and then shrinks in size when the rotation rate ratio increases till $\mu = 0.5$.}
As $\mu$ is increased further, the island seems to again stretch in the direction of the Taylor-number axis and shrink on the axial--wavenumber axis. 
We could continue the island till $\mu = 0.95$ and at $\mu = 1$, the left and right boundaries separate to form an (seemingly) open the neutral stability curve. This explains why we could not find CDNC for $m = -4$ mode in the wide--gap at $\mu = 1$.

Finally, Fig. \ref{fig:island_origin_of_small_k_instab} \aghoredit{(b)} shows the variation of an instability island for the thin--gap annulus that emerges at $\mu = 0.6$ through $\mu = 1$. The sequence of variation of the CDNC in the thin--gap is much like the one shown in Fig. \ref{fig:island_origin_of_small_k_instab} \aghoredit{(a)} for the wide--gap case, just at different values of the rotation rate ratio. It can be seen that the instability island occupies more area in the Taylor number versus axial perturbation wavenumber space as $\mu$ increases from $0.6$ to $0.7$. From $\mu = 0.7$ to $\mu = 0.95$, the island consistently moves towards smaller values of $k_z$ before turning into an (seemingly) open neutral stability boundary in the pure co--rotation case $\mu = 1$.

\begin{figure}
    \centering
    \includegraphics[scale=0.49]{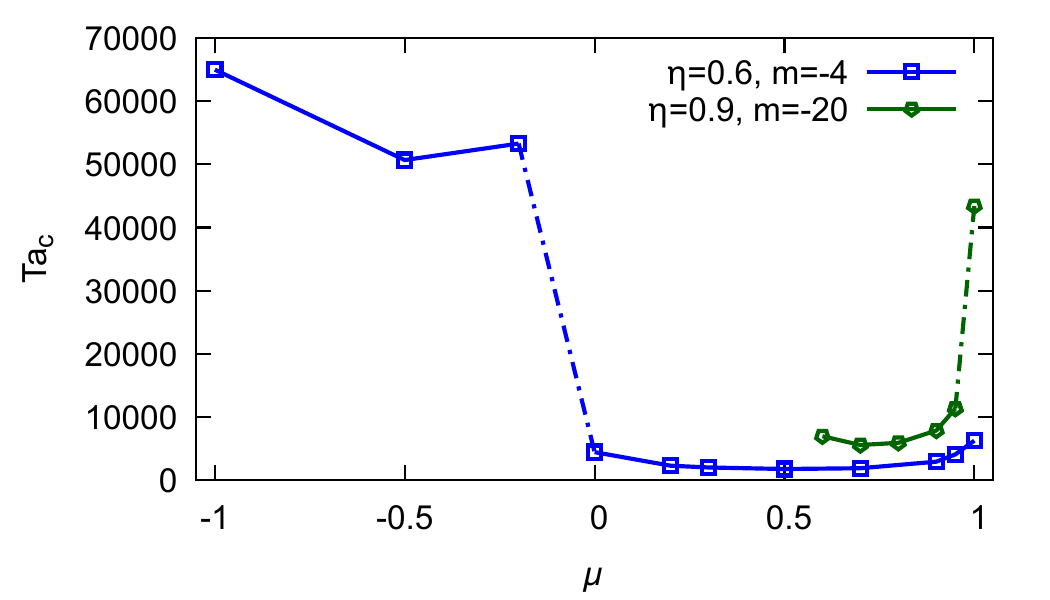}
    \caption{\aghoredit{Discontinuous jumps in the critical Taylor numbers induced by appearance/disappearance and morphing of instability islands. The critical Taylor numbers for the azimuthal modes $m = -4$ in the wide gap case and $m = -20$ in the thin gap case are plotted as a function of rotation rate ratios. }}
    \label{fig:tac-jump}
\end{figure}

Figures \ref{fig:island_eta_0.6_m_-3} and \ref{fig:island_origin_of_small_k_instab} collectively demonstrate that CDNCs are (i) sensitive to changes in the rotation rate ratio, (ii) can spilt from, merge into or morph into open neutral stability curves or (iii) disappear as the rotation rate ratio is varied.
\aghoredit{The CDNCs are also responsible for discontinuous jumps in the critical Taylor number $Ta_c$. 
This is illustrated in Fig. \ref{fig:tac-jump} where it can be seen that for $m=-4$ in the wide gap, the critical Taylor number reduces discontinuously as the rotation rate ratio is changed from $-0.2$ to $0.0$.
Similarly, for $m=-20$ in the thin gap, the critical Taylor number increases discontinuously as the rotation rate ratio is changed from $0.95$ to $1.0$.
The discontinuous jump shown in Fig. \ref{fig:tac-jump} for the wide gap case (with $m = -4$) is associated with the appearance of an instability island at $\mu = 0$. 
On the other hand, the discontinuous jump for the thin gap (with $m = -20$) is associated with the morphing of the instability island into an open neutral curve at $\mu = 1$, see Fig. \ref{fig:island_origin_of_small_k_instab}(b). 
Similar morphing occurs in the wide gap (with $m = -4$) when the rotation rate ratio is increased from $\mu = 0.95$ to $1$, see Fig. \ref{fig:island_origin_of_small_k_instab} (a). However, no discontinuous jump is observed in Fig. \ref{fig:tac-jump} for the wide gap case when the rotation rate ratio is increased from $\mu = 0.95$ to $\mu = 1$. With these observations we conclude that appearance/disappearance of a CDNC always induces a discontinuity in the critical Taylor number, while morphing of CDNCs into open curves might or might not induce a discontinuous jump in the critical Taylor number. }


\section{Discussion}
\label{sec:5}
We now discuss the observations of our analysis and their implications.
Here, we focus on the fastest growing modes and the effect of outer cylinder rotation and radius ratio on the linear instability thresholds. 

\subsection{Qualitative description of the fastest growing modes}
In all cases reported in this article, neutral modes at the onset of instability are found to be oscillatory, with critical frequencies reported in Table \ref{tbl:Gr_1000_neutral}. 
It is seen from the table that depending on the parameters $\{\eta, \mu\}$, either of axisymmetric toroidal modes ($m = 0$) or non-axisymmetric spiral modes can grow at the fastest rate.
The fastest growing modes at the onset of instability are seen to be confined near the inner cylinder for $\mu = -0.5, -1$ for both wide--and thin--gap cases (Figs. \ref{fig:most_unstable_eta_0.6_G_1000_counterrot},\ref{fig:most_unstable_eta_0.9_Gr_1000_counterrot}). 
This effect is more pronounced in the wide--gap case ($\eta = 0.6$) evident from the $2d$ slices of the reconstructed $3d$ critical eigenmodes. 
Thus, we can conclude that the specific angular momentum $(\mathcal{L} = v r)$ transport from the inner wall to the outer wall is suppressed for $\mu \in [-0.5, -0.1]$ with increasing counter--rotation. 

\aghoredit{One of the important results of this paper was the discovery of unstable modes in the Rayleigh--stable regime (Fig. \ref{fig:neutral_stab_G_1000_eta_0.6} and Fig. \ref{fig:neutral_stab_G_1000_eta_0.9}  (d), for example). The modes in the pure co--rotation case ($\mu = 1$) are presented and discussed in Figs. \ref{fig:most_unstable_eta_0.6_G_1000_corot} and \ref{fig:most_unstable_eta_0.9_G_1000_corot} for both wide and thin gaps. These modes are due to buoyancy, since in the isothermal case, these modes would not exist according to the classical Rayleigh--criterion. This implies the existence of a critical Grashof number $0 < \Gr_c \leq 1000$ after which buoyancy effects become important. This task might be taken up in a future work.} 

For the wide--gap case with $\mu = 0.5$, we find closed islands of instability for azimuthal modes $m = -3$ and $-4$. 
For the pure co--rotation case, a seemingly open neutral stability curve was found for $m= -4$ and no unstable region was found for $m = -3$. 
In the thin--gap case, we find $m = -20$ to be the fastest growing mode with $m = -19, -21$ having close critical Taylor numbers $\Ta_c$ for the onset of instability in the pure co--rotation case. 

\subsection{Symmetry--breaking and bifurcation theory}\label{sec:symm_breaking}
The isothermal TCF is O(2) symmetric in the axial co--ordinate $z$ (invariant under reflections and translations), i.e., if $\bu(r, \phi, z, t)$ is a solution then so are $\bu(r, \phi, -z, t)$ and $\bu({r, \phi, z-z_0, t})$ for any $z_0 \in \bbR$. 
For the O(2) symmetry, Hopf bifurcations are degenerate and symmetry--breaking, i.e., bidirectional traveling waves simultaneously bifurcate and have the same speeds. 
Linear analysis cannot comment on the nature of the bifurcating solution in the degenerate Hopf bifurcation (see \citet{golubitsky2012singularities} for more details). 
Thus, whether the solution takes form of a traveling wave or a standing wave (an equal superposition of two traveling waves traveling in the opposite direction at the same speed) is determined by nonlinear analysis \cite{graham1998effect}.
In the case of radially heated TCF, the reflection symmetry is broken by the axial base flow and the flow is SO(2) symmetric, retaining only the translation symmetry in the axial co--ordinate. In such a case, Hopf bifurcations are non--degenerate and linear analysis {is} sufficient to predict the form of the bifurcating solution \cite{iooss1990bifurcation,crawford1988degenerate}. 
In the present problem we observe that critical modes with $m \geq 0$ have a negative $\sigma_{ic}$ while $m < 0$ have a positive $\sigma_{ic}$. 
Therefore, all the critical non-axisymmetric helical modes with $m \geq 0$ travel with an upward axial speed, while $m < 0$ travel with a downward axial speed according to Eq. \eqref{eq:perturbation_shape}.

Furthermore, we observe that the modes $\pm m$ for a given set of control parameters $\{\eta, \Gr, \mu\}$ do not become unstable simultaneously and the speeds of the solutions bifurcating from the basic state are also different. 
Since $\pm m$ modes become unstable at different Taylor numbers, we find both positive and negative $m$ to be the fastest growing modes depending on parameters, indicating negative and positive inclination $\psi$, respectively, of the lines of constant phase with respect to the horizontal direction according to Eq. \eqref{eq:perturbation_shape}. 
This behavior might be explained with a perturbed normal form for the O(2)--symmetric Hopf bifurcation with broken reflection symmetry, see \citet{abshagen2007imperfect} for example. 
\subsection{Effect of varying rotation rate ratio and radius ratio on the fastest growing modes}

We now shift the focus of our discussion from qualitative to quantitative and describe the effects of varying one of $\{\eta, \mu \}$ while the other control parameter is held constant.
 \begin{figure*}
 \begin{center}
  \includegraphics[width = 0.98\textwidth]{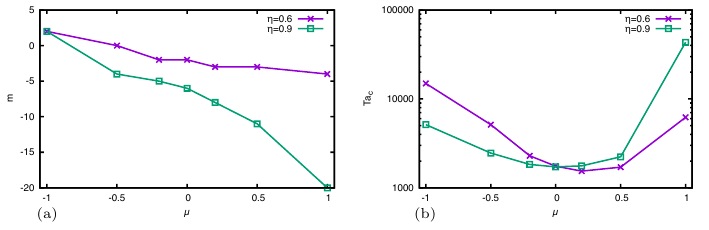}
\end{center}
\caption{Quantities plotted here correspond to the highlighted entries in Table \ref{tbl:Gr_1000_neutral}. \aghoredit{(a)} Azimuthal wavenumber $m$ for the fastest growing mode versus rotation rate ratio $\mu$. As $\mu$ is increased higher azimuthal modes are seen to become unstable first.  \aghoredit{(b)} A semi-log plot of the critical Taylor number $Ta_c$ versus $\mu$. The plot highlights the general stabilizing trend when the outer cylinder is rotated, with an exception. In the wide--gap case, when $\mu$ is increased from  $0$ to $0.2$, the stabilizing effect of rotating outer cylinder is observed, reflected by a dip in the critical Taylor number. Refer to text for more discussion.}
 \label{fig:s}
\end{figure*}
First, we fix $\eta $ and observe the effect of changing $\mu$. 
This amounts to comparing the highlighted values, moving in the vertical direction in Table \ref{tbl:Gr_1000_neutral} to see the effect of changing $\mu$ for a fixed $\eta$. 
Fig. \ref{fig:s} \aghoredit{(a)} plots the azimuthal wavenumber $m$ for the critical modes as the rotation rate ratio $\mu$ is varied. 
As $\mu$ increases from pure counter--rotation through pure co--rotation, higher azimuthal successively higher wave numbers become unstable first (evidenced by increased absolute value of $m$).  
Fig. \ref{fig:s} \aghoredit{(b)} plots $\Ta_c$ corresponding to the fastest growing mode for different values of $\mu$ at a given radius ratio. 
In general, $Ta_c$ for $\mu \neq 0$ is larger in comparison to $Ta_c$ for $\mu = 0$, indicating a general stabilizing effect. 
In the case $\eta = 0.6$, counter--rotation stabilizes the flow (as $\mu$ is lowered below $0$, $Ta_c$ increases in comparison to when $\mu = 0$). 
However co--rotation has an initial destabilizing effect, shown by the decrease in $Ta_c$ when $\mu$ is increased from $0$ to $0.2$, followed by a stabilizing effect. 

Next, we compare critical Taylor numbers at $\eta = 0.6$ and $0.9$ for a constant $\mu$ to assess the impact of increasing radius ratio. 
This amounts to comparing highlighted values across columns of different $\eta$ for a fixed value of $\mu$. 
Alternatively, it is equivalent to comparing the two curves for a fixed value $\mu$ in Fig. \ref{fig:s}. 
From Fig. \ref{fig:s} \aghoredit{(a)}, it is clear that the azimuthal wavenumbers of the critical modes are in general larger in magnitude for the thin-gap case as compared to the wide-gap case. 
From Fig. \ref{fig:s} \aghoredit{(b)} it is seen that the critical Taylor numbers for the thin--gap case are lower than those for the wide--gap case for $\mu \leq 0$, thus suggesting that increasing $\eta$ has a destabilizing effect in the Rayleigh-unstable counter--rotation regime.
For $\mu = 0.2$, a stabilizing effect of increasing $\eta$ (evidenced by increased $Ta_c$).
The case of $\mu = 0.5$ is qualitatively different for the two radius ratios, i.e., it is in the Rayleigh-stable regime for $\eta = 0.6$, but in the Rayleigh-unstable regime for $\eta = 0.9$. 
Lastly, the co--rotation case ($\mu = 1$) at $\Gr=1000$, although lying in the Rayleigh-stable regime, is found to be unstable at $Ta \sim 6 \times 10^{3}$ at $\eta = 0.6$, while at $\eta = 0.9$ is found to be unstable at $Ta \sim 5 \times 10^{4}$, thus implying a stabilizing effect of increasing $\eta$ for the co--rotation case in the Rayleigh-stable regime. 

For both the radius ratios considered, CDNCs/instability islands are found in the parameter space that show restabilization with increasing Taylor numbers. 
The CDNCs are sensitive to changes in the control parameter $\mu$ as can be seen from the varied range of Taylor numbers and the axial perturbation wavenumbers $k_z$ in Figs. \ref{fig:island_eta_0.6_m_-3} and \ref{fig:island_origin_of_small_k_instab}. 
Their disappearance can significantly affect the critical Taylor number for the onset of instability. 
For example, we found an instability island for $m = -4, \mu = 0$, but could not find it for $m =-4, \mu = -0.2$. 
The discontinuous jump in the critical Taylor number ($Ta = 53308.15$ for $\mu = -0.2$ and $Ta = 4390$ for $\mu = 0$) and axial wavenumber of the perturbation ($k_z = 7.42720$ for $\mu = -0.2$ and $k_z = 4.1040$ for $\mu = 0$) observed for mode $m = -4$ (see Table \ref{tbl:Gr_1000_neutral}) can be thus attributed to the appearance of the instability island as $\mu$ is increased from $-0.2$ to $0$. 
Moreover, in all cases of islands considered here, the point of onset of instability is seen to consistently move towards smaller axial perturbation wavenumbers as $\mu$ is increased. 
Therefore, longer--wavelength instabilities are set up as rotation rate ratio is increased from $\mu=0$ for the pure inner cylinder rotation to $\mu=1$ in the pure co--rotation case. 

\section{Conclusions and future work}
\label{sec:6}
We conducted a systematic linear stability analysis of the radially--heated Taylor--Couette flow and observed the effects of outer--cylinder rotation.
All calculations were performed at a fixed value of Grashof number $\Gr = 1000$ and Prandtl number $Pr = 1$. 
Two radius ratios $\eta \in \{0.6, 0.9\}$, representing wide--gap and thin--gap cases were considered. 
For each radius ratio, intermediate rotation rate ratios were considered ranging from pure counter--rotation ($\mu = -1$) to pure co--rotation ($\mu = 1$).
We focused on the small Froude number limit $\Fr \rightarrow 0$ where centrifugal buoyancy can be neglected. 
In this limit, we extended symmetries of the linearized system identified by \citet{ali1990stability} to include nonzero $\mu$ as defined in Eqs. \eqref{eq:symms1}, \eqref{eq:symms2}, and \eqref{eq:S3}. 
We also provided possible explanations for the observed behavior using a combination of symmetry arguments and bifurcation theory. 
Weakly nonlinear analysis via amplitude equations would provide further insights in the bifurcation scenario for the radially heated TCF and this is left as future work.  

Closed regions of instability in the Taylor number versus axial perturbation wavenumber plane for both wide-- and thin--gaps were located. 
The origin and disappearance of CDNCs was investigated for some important azimuthal modes. 
CDNCs were shown to be sensitive to changes in $\mu$. 
A sweep over intermediate values of $\mu$ showed how an island can separate from or merge into an open neutral stability boundary, for example, $m = -3$ mode in the wide--gap case, see Fig. \ref{fig:island_eta_0.6_m_-3}, $\mu = 0.3$ to $0.31$. 
Alternatively, the island can itself morph into an open neutral stability boundary (see Fig. \ref{fig:island_origin_of_small_k_instab}, $\mu = 0.95$ to $1$). 
The latter case revealed the origin of the (seemingly) open neutral stability boundaries found at small $k_z$ in the pure co--rotation case and showed that they can be traced back to CDNCs as rotation rate ratio $\mu$ is decreased, for example, see Fig. \ref{fig:island_origin_of_small_k_instab}. 

An important result of this paper is that the appearance/disappearance of instability islands can significantly change the critical Taylor number for onset of instability. 
For $m = -3$ in the wide--gap case, we could continue the island born at $\mu = 0.31$ till $\mu = 0.93$ and could not find an unstable mode for $\mu \geq 0.95$. Similarly, the island found at $\mu = 0.5$ for mode $m= -4$ in the wide--gap case could be continued down in $\mu$ till $\mu = 0$. Therefore, sudden discontinuities in the critical Taylor number can be explained via appearance/disappearance of the instability islands. 
If such an island is located, restabilization of the flow with increasing Taylor numbers can also be observed experimentally.

The analysis presented here can be extended to examine the effects of varying Prandtl numbers. It would be especially interesting to study moderate Prandtl numbers where shear and buoyancy effects compete. Another interesting avenue would be to investigate the effect of centrifugal buoyancy (non-vanishing Froude numbers) when outer-cylinder rotation is non-zero ($\mu \neq 0$).  
Linear stability analysis also provides insight into an interesting range of parameters for performing direct numerical simulations of the governing equations which could shed light on the rich nonlinear dynamics of the system.


\appendix

\section{Table of critical values}

\begin{center}
\setlength\LTleft{0pt}
\setlength\LTright{0pt}
\setlength{\LTcapwidth}{\textwidth}
\begin{longtable*}{|cc|ccc|ccc|}
    \toprule
    \multirow{1}{*}{$\mu$} & 
    \multirow{1}{*}{$m$} &
    \multicolumn{3}{c}{$\eta = 0.6$} &
    \multicolumn{3}{c}{$\eta = 0.9$} \\
       &  & 
      $Ta_c$ & $k_c$ & $\sigma_{ic}$ & $Ta_c$ & $k_c$ & $\sigma_{ic}$\\
      \hline
       -1 & -5 &   - & - & -                         & 5822.94 & 3.79609 & 2.66661e+01 \\
       -1 & -4 &   - & - & -                         & 5904.26 & 3.58278 & 2.06109e+01 \\
       -1 & -3 &  23060.40 & 5.93754 & 8.7977e+01    & 6232.08 & 4.16401 & 9.10370e+00 \\
       -1 & -2 &  18572.05 & 5.85807 & 4.61341e+01   & 6213.97 & 4.47134 & -4.03754e+00 \\
       -1 & -1 &  17780.14 & 6.15256 & 7.72213e-01   & 5975.70 & 4.44946 & -1.52341e+01 \\
       -1 & 0  &  17092.65 & 6.23491 & -4.59031e+01  & 5558.93 & 4.19639 & -2.00041e+00 \\
       -1 & 1  &  15823.30 & 5.83604 & -8.42856e+01  & 5165.36 & 3.85050  &  -2.94438e+01 \\
       -1 & 2  & \textbf{14933.89} & \textbf{4.86187} & \textbf{-1.94001e+01} & \textbf{5138.40} & \textbf{3.68113}  & \textbf{-3.42262e+01} \\
       -1 & 3 & 27408.38 & 6.89704 & -1.94146e+02   & 5622.58 & 3.70697 & -4.08248e+01 \\
       -1 & 4 & 54537.12 & 9.22380 & -4.03154e+02 & 6773.24 & 3.92187 & -5.20952e+01 \\
      \\      
      -0.5 & -5 &  - & - & -   & 2545.10 & 3.34599 & 2.10345e+01 \\
      -0.5 & -4 & 50694.13 & 7.84844 & 3.1524e+02   & \textbf{2460.55} & \textbf{3.18859} & \textbf{1.51461e+01} \\
      -0.5 & -3 & 9814.35 & 4.89194 & 5.7900e+01   & 2476.22 & 3.11815 & 9.64734e+00 \\
      -0.5 & -2 & 5875.06 & 3.69844 & 2.3731e+01   & 2568.18 & 3.14012 & 3.88648e+00 \\
      -0.5 & -1 & 5701.57 & 4.05867 & -6.7819e-01   & 2714.67 & 3.21703 & -2.42134e+00 \\
      -0.5 & 0  & \textbf{5135.82} & \textbf{3.88670} & \textbf{-2.2838e+01}    & 2906.45 & 3.28911 & -9.21665e+00 \\
      -0.5 & 1  & 5713.14 & 3.70639 & -3.95978e+01   & 3161.99 & 3.33462 & -1.64688e+01 \\
      -0.5 & 2  & 10030.80 & 4.32672 & -7.83492e+01   & 3520.28 & 3.36384 &  -2.44289e+01 \\
       -0.5 & 3 & 22210.20 & 6.36873 & -2.01605e+02  & - & - & - \\
      \\
      -0.2 & -6  & - & - & - & 1854.91 & 3.10887 & 3.26209e+01  \\
      -0.2 & -5  & - & - & - &  \textbf{1832.02} & \textbf{2.99341} & \textbf{2.64812e+01}  \\
      -0.2 & -4  & 53308.15 & 7.42720 & 3.8069e+02 & 1864.91 & 2.92531 & 2.06957e+01 \\
      -0.2 & -3  & 3338.60 & 3.64635 & 3.8570e+01  & 1946.64 &  2.90953 & 1.49354e+01 \\
      -0.2 & -2  &  \textbf{2292.02} & \textbf{2.87532} & \textbf{1.8710e+01} & 2075.38 & 2.93694 & 8.88949e+01 \\
       -0.2 & -1 & 2497.02 & 2.80735 & 5.9476e+00  & 2250.29 & 2.99935 & 2.27335e+00 \\
      -0.2 & 0 & 3287.03 & 3.20385 & -1.0691e+01 & 2476.38 & 3.07785 & -5.09879e+00 \\
      -0.2 & 1 & 4815.42 & 3.44300 & -3.33710e+01 & 2758.99 & 3.15405 & -1.33962e+01 \\
      \\
      0 & -7  & - & - & -                       & 1728.44  & 2.99275 & 4.67167e+01 \\
      0 & -6  & - & - & -                       & \textbf{1719.71}  & \textbf{2.89803} & \textbf{3.96288e+01} \\
      0 & -5  & - & - & -                       & 1751.73 & 2.84231 & 3.29572e+01  \\
      0 & -4  & 4390 & 4.1040 & 8.6151e+01 & 1818.71 & 2.82611 & 2.63673e+01 \\
      0 & -3  & 1863.57 & 2.75454 & 4.0911e+01  & 1920.80 & 2.84045 & 1.96343e+01 \\
      0 & -2  & \textbf{1750.89} & \textbf{2.40775} & \textbf{2.5263e+01}  & 2061.02 & 2.88257 & 1.25388e+01 \\
      0 & -1 & 2279.48 & 2.52391 & 1.1727e+01   & 2238.61 & 2.94673 & 4.84410e+00 \\
      0 & 0  & 3413.10 & 3.05819 & -7.5626e+00  & 2461.40 & 3.01898 & -3.62633e+00 \\
      0 & 1  & 5447.53 & 3.36079 & -3.63104e+01 & 2738.91 & 3.08702 & -1.30737e+01 \\
      0 & 2  & - & - & -                        & 3083.91 & 3.14313 & -2.37368e+01 \\
      0 & 3  & - & - & -                        & 3531.08 & 3.19391 & -3.60056e+01 \\
      \\ 
      0.2 & -9 & - & - & -                        & 1791.74 & 2.95175 & 7.48679e+01  \\
      0.2 & -8 & - & - & -                        & \textbf{1765.43} & \textbf{2.85565} & \textbf{6.58774e+01}  \\
      0.2 & -7 & - & - & -                        & 1773.26 & 2.78668 & 5.75972e+01  \\
      0.2 & -6 & - & - & -                        & 1810.61 & 2.74096 & 4.96448e+01  \\
      0.2 & -5 & - & - & -                        & 1879.73 & 2.73007 & 4.18579e+01  \\
       0.2 & -4 & 2243.00 & 2.94213 & 8.2493e+01 & 1978.18 & 2.74627 & 3.39657e+01 \\
      0.2 & -3 & \textbf{1541.34} & \textbf{2.18362} & \textbf{5.1182e+01} & 2106.27 & 2.78474 & 2.57563e+01 \\
      0.2 & -2 & 1781.54 & 1.99982 & 3.5861e+01   & 2268.16 & 2.84706 & 1.70398e+01 \\
      0.2 & -1 & 2971.26 & 2.16111 & 2.1219e+01   & 2463.15 & 2.91645 & 7.62532e+00 \\
      0.2 & 0  & 5655.64 & 3.00812 & -6.0502e+00  & 2697.45 & 2.99092 & -2.66574e+00 \\
      0.2 & 1  & - & - & -                        & 2983.03 & 3.05846 & -1.4019e+01 \\
      0.2 & 2  & - & - & -                        & 3324.88 & 3.11715 & -2.66710e+01 \\
      \\   
      0.5 & -12 & - & - & -                       & 2264.83 & 2.67458 & 1.43227e+02  \\
      0.5 & -11 & - & - & -                       & \textbf{2232.38} & \textbf{2.57911} & \textbf{1.30267e+02}  \\
      0.5 & -10 & - & - & -                       & 2235.43 & 2.52124 & 1.18428e+02  \\
      0.5 & -9 & - & - & -                        & 2273.36 & 2.48403 & 1.07387e+02  \\
      0.5 & -8 & - & - & -                        & 2342.76 & 2.46297 & 9.68018e+01  \\
      0.5 & -7 & - & - & -                        & 2447.51 & 2.45833 & 8.64365e+01  \\
      0.5 & -6 & - & - & -                        & 2586.92 & 2.47868 & 7.60139e+01  \\
      0.5 & -5 & - & - & -                        & 2760.59 & 2.52177 & 6.52176e+01  \\
      0.5 & -4 & 1714.98 & 1.72996 & 1.0341e+02   & 2970.68 &  2.59625 & 5.38514e+01 \\
      0.5 & -3 & \textbf{1706.85} & \textbf{1.50802} & \textbf{7.7196e+01} & 3215.85 & 2.68364 & 4.16453e+01 \\
      0.5 & -2 & $\NA$ & $\NA$ & $\NA$            & 3497.51 & 2.78294 & 2.84293e+01 \\
      0.5 & -1 & $\NA$ & $\NA$ & $\NA$            & 3818.61 & 2.88102 & 1.40207e+01 \\
      0.5 & 0  & $\NA$ & $\NA$ & $\NA$            & 4182.63 & 2.97346 & -1.73913e+00 \\
      0.5 & 1  & $\NA$ & $\NA$ & $\NA$            & 4593.67 & 3.05210 & -1.90168e+01 \\
      0.5 & 2  & - & - & -                        & 5083.80  & 3.11946 & -3.80876e+01 \\
      0.5 & 3  & - & - & -                        & 5643.29 & 3.16986 & -5.91703e+01 \\
      0.5 & 4  & - & - & -                        & 6340.51 & 3.20545 & -8.29205e+01 \\
      \\
      1 & -22 & - & - & -                        & 48851.97 & 0.583363 & 1.66520e+03 \\
      1 & -21 & - & - & -                        & 45648.53 & 0.58320 & 1.53650e+03\\
      1 & -20 & - & - & -                        & \textbf{43411.00} & \textbf{0.58277} & \textbf{1.42701e+03} \\
      1 & -19 & - & - & -                        & 44753.92 & 0.58333 & 1.37646e+03\\
      1 & -18 & - & - & -                        & 48272.25 & 0.53983 & 1.35431e+03\\
      1 & -4 & \textbf{6209.99} & \textbf{0.56951} & \textbf{2.9684e+02} & - & - & -\\
      1 & -3 & $\NA$ & $\NA$ & $\NA$ & - & - & -\\
      1 & -2 & $\NA$ & $\NA$ & $\NA$ & - & - & -\\
      1 & -1 & $\NA$ & $\NA$ & $\NA$ & - & - & -\\
      1 & 0 & $\NA$ & $\NA$ & $\NA$  & - & - & -\\
      1 & 1 & $\NA$ & $\NA$ & $\NA$  & - & - & -\\
    \bottomrule
  \caption{Critical values for $\eta = 0.6,0.9$. The `$\NA$' entries represent cases where the reported modes are found to be stable up to a Taylor number of $10^{5}$ and entries with `-' represent cases where the fastest growing mode had already been located. Highlighted values correspond to the fastest growing modes.}
  \label{tbl:Gr_1000_neutral}
\end{longtable*}
\end{center}

\nocite{*}
\bibliography{main}

\end{document}